\documentclass[sigconf]{acmart}

\AtBeginDocument{%
  \providecommand\BibTeX{{%
    Bib\TeX}}}

\setcopyright{acmlicensed}
\copyrightyear{2018}
\acmYear{2018}
\acmDOI{XXXXXXX.XXXXXXX}
\acmConference[Conference acronym 'XX]{Make sure to enter the correct
  conference title from your rights confirmation email}{June 03--05,
  2018}{Woodstock, NY}
\acmISBN{978-1-4503-XXXX-X/2018/06}

\usepackage{multirow}

\begin{document}

\title{Adaptive User Interest Modeling via Conditioned Denoising Diffusion For Click-Through Rate Prediction}


\author{Qihang Zhao}
\email{zhaoqh75@mail.ustc.edu.cn}
\affiliation{%
  \institution{Kuaishou Inc.}
  \city{Hangzhou}
  \state{Zhejiang}
  \country{China}
}

\author{Xiaoyang Zheng}
\affiliation{%
  \institution{Kuaishou Technology}
  \city{Hangzhou}
  \country{China}}
\email{zhengxiaoyang@kuaishou.com}

\author{Ben Chen}
\affiliation{%
  \institution{Kuaishou Technology}
  \city{Hangzhou}
  \country{China}}
\email{chenben03@kuaishou.com}

\author{Zhongbo Sun}
\authornote{*corresponding author}
\affiliation{%
  \institution{Kuaishou Technology}
  \city{Hangzhou}
  \country{China}}
\email{sunzb17@gmail.com}

\author{Chenyi Lei}
\affiliation{%
  \institution{Kuaishou Technology}
  \city{Hangzhou}
  \country{China}}
\email{leichy@mail.ustc.edu.cn}





\renewcommand{\shortauthors}{Zhao et al.}

\begin{abstract}
User behavior sequences in search systems resemble "interest fossils"—capturing genuine intent yet eroded by exposure bias, category drift, and contextual noise. Current methods predominantly follow an "identify-aggregate" paradigm, assuming sequences immutably reflect user preferences while overlooking the organic entanglement of noise and genuine interest; moreover, they output static, context-agnostic representations, failing to adapt to dynamic intent shifts under varying Query-User-Item-Context conditions.

To resolve this dual challenge, we propose the Contextual Diffusion Purifier (CDP). By treating category-filtered behaviors as "contaminated observations", CDP employs a forward noising and conditional reverse denoising process guided by cross-interaction features (Query × User × Item × Context), controllably generating pure, context-aware interest representations that dynamically evolve with scenarios. Extensive offline/online experiments demonstrate the superiority of CDP over state-of-the-art methods.

\end{abstract}


\received{20 February 2007}
\received[revised]{12 March 2009}
\received[accepted]{5 June 2009}

\maketitle

\section{Introduction}
Amid the explosive growth of internet content, search and recommendation systems have become primary gateways for users to access information, products, and services. Behind each query, platforms must infer genuine user intent within millisecond-level latency and deliver optimally ranked results. To achieve this, systems rely heavily on historical behavior sequences—spanning clicks, views, favorites, purchases, and other multidimensional signals—as core evidence for profiling user interests. These sequences not only reflect past choices but also implicitly encode future intent, making them critical for enhancing ranking efficacy in both industry and academia .

However, real-world user behaviors are never pure expressions of interest. They emerge from the interplay of exposure mechanisms, position bias, category drift, erroneous clicks, and real-time context, resulting in sequences that entangle genuine preferences with substantial environmental noise. This coupling obscures signal separation and compromises representation fidelity .

To distill actionable signals from behavioral data, three methodological paradigms have emerged: Early aggregation models compress sequences into fixed-length vectors via average or max pooling. While computationally efficient and deployment-friendly, they ignore behavioral heterogeneity and contextual dynamics . Attention-based models (e.g., DIN~\cite{zhou2018deep}, DIEN~\cite{10.1609/aaai.v33i01.33015941}, SASRec ~\cite{10.1145/3695719.3695727}) weight behaviors differently to highlight critical signals. Yet, their attention mechanisms—supervised solely on raw observations—amplify biases in noisy scenarios, as they lack explicit noise-disentangling capability . Generative approaches (e.g., HSTU~\cite{10.5555/3692070.3694484}, MTGR~\cite{han2025mtgr}) predict next behaviors via autoregressive or masked tasks. Despite improved generalization, their objectives neglect explicit modeling of contextual triggers (e.g., queries, items, real-time settings) that dynamically reshape interests .
Critically, all existing methods operate under the assumption that observed behaviors equate to ground truth, failing to decouple noise from preferences or adapt representations to shifting contexts .

To resolve these dual limitations, we propose the Contextual Diffusion Purifier (CDP). Our framework first filters out category-inconsistent behaviors to isolate a category-aligned subsequence, mitigating cross-category noise. This subsequence is treated as a contaminated observation. Guided by a conditional quadrangle (Query × User × Item × Context), a diffusion-based denoising process then generates a pure, context-adaptive interest representation. Crucially, CDP integrates seamlessly into existing ranking architectures without structural modifications, ensuring scalability and interpretability .

Our key contributions are:
\begin{itemize}
  \item We propose a novel framework for modeling user behavior sequences: guided by context, conditionally controlled to generate user adaptive historical behavioral interests.
  
  \item We design a multi condition fusion mechanism based on MoE to adaptively extract appropriate condition signals and achieve personalized condition guidance.
  
  \item Extensive offline and online experiments reveal the effec-
tiveness of our proposed method.
\end{itemize}

\section{Related Work}

\subsection{User Historical Behavior Sequence Modeling}

The evolution of user-behavior sequence modeling has been marked by a progressive decoupling of retrieval efficiency and representation fidelity. SIM~\cite{10.1145/3340531.3412744} introduced a two-stage architecture with a coarse-grained retrieval module followed by a fine-grained attention module. ETA~\cite{chen2021end} replaced the heuristic retrieval criterion with a learnable locality-sensitive hashing mechanism. QIN~\cite{10.1145/3583780.3615022} extended this pipeline by integrating query-aware representations into the retrieval key. To eliminate the objective discrepancy between retrieval and ranking stages, TWIN~\cite{10.1145/3580305.3599922} proposed a consistency-preserved retrieval unit that shares the exact relevance metric employed by the succeeding attention layer, while TWIN\_v2~\cite{10.1145/3627673.3680030} further advanced this approach by compressing life-cycle sequences exceeding ten million interactions through hierarchical adaptive clustering, preserving long-term semantics within constant memory footprints. ADS~\cite{10.1145/3726302.3731939} generalized the paradigm to heterogeneous traffic by introducing domain-adaptive scaling factors that modulate attention magnitudes per segment, ensuring robust transfer across different platforms without re-training. MARM~\cite{DBLP:journals/corr/abs-2411-09425} augmented the pipeline with a KV-cache-style memory layer for ultra-long sequences. HSTU~\cite{10.5555/3692070.3694484} represented a more radical departure by abandoning the retrieval-attention dichotomy and reformulating sequential recommendation as a generative transduction task within a single trillion-parameter sparse-temporal backbone. Finally, LONGER~\cite{chai2025longer} introduced elastic token-merge transformers and GPU-friendly key–value caches, enabling synchronous dense-plus-sparse training on sequences exceeding ten million tokens and marking the transition from retrieval-attention coupling to fully generative lifelong modeling.

\subsection{Diffusion Model for Recommendation}
Diffusion probabilistic models (DPMs) have evolved into a powerful framework for building robust and controllable recommender systems. CaDiRec~\cite{10.1145/3627673.3679655} enhances sequential recommendation by integrating user-contextual information into the diffusion process and introducing contrastive regularization. CDDRec~\cite{10.1007/978-981-97-2262-4_13} treats the user’s historical behavior sequence as a conditional input, guiding the diffusion model to generate recommendation sequences that align with evolving interests during the denoising process. Diff4Rec~\cite{10.1145/3581783.3612709} employs a curriculum-scheduled diffusion-augmentation mechanism in the latent space to generate augmented sequences, effectively alleviating data sparsity in sequential recommendation. DDRM~\cite{10.1145/3626772.3657825} iteratively denoises user/item embeddings by leveraging collaborative signals as guidance and personalized historical embeddings as starting points, enhancing robustness against noisy implicit feedback. DiffRec~\cite{10.1145/3539618.3591663} introduces a denoising-diffusion recommender that recovers clean user interactions from minimal-noise perturbations, its extensions L-DiffRec(latent-space diffusion for scalability) and T-DiffRec(time-aware reweighting for dynamic preferences) further address practical challenges in large-scale and sequential scenarios.

\section{Methodology}
\label{sec:method}

\begin{figure*}
    \centering
    \includegraphics[width=0.96\textwidth]{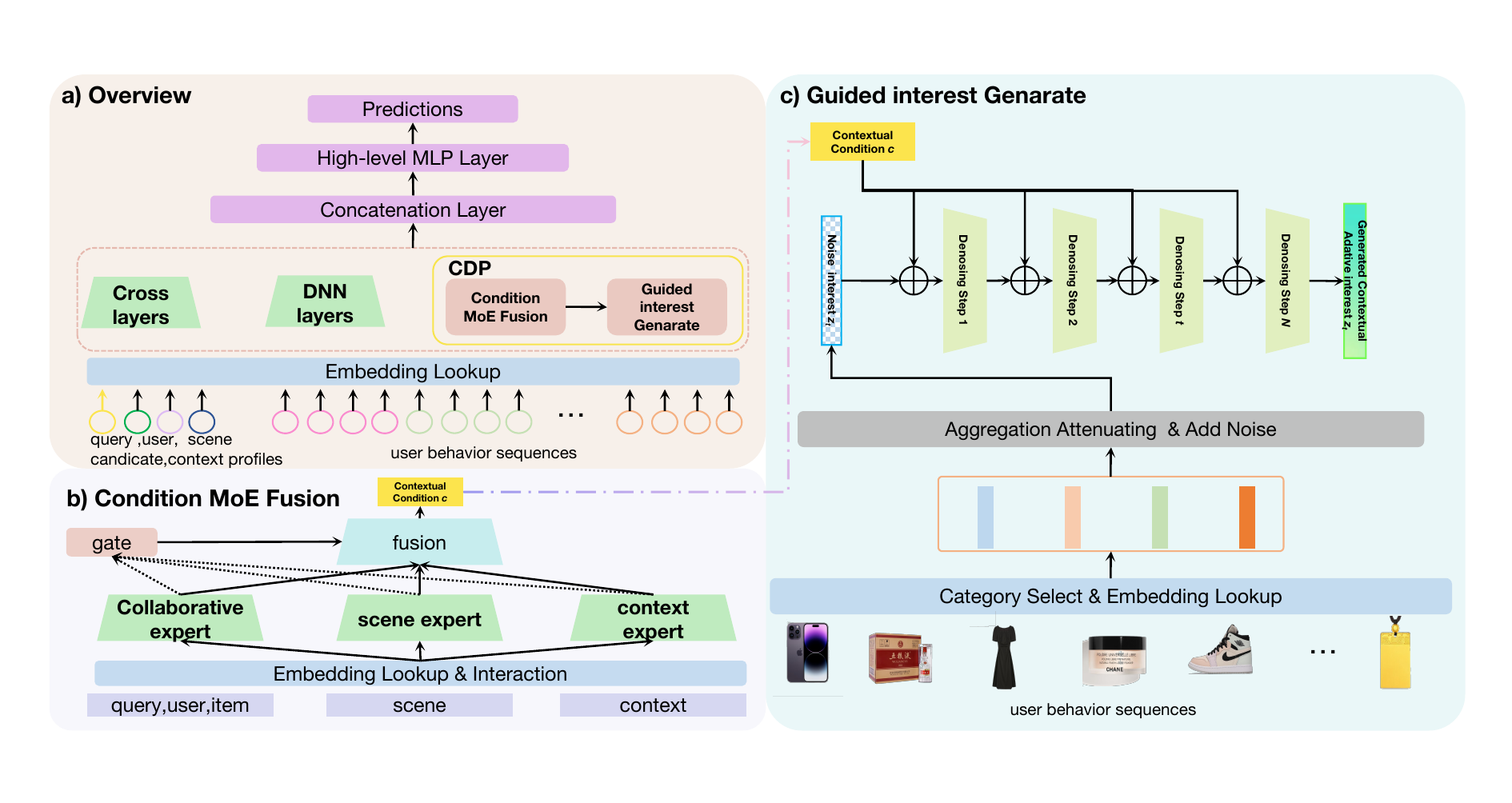}
    \caption{Overall framework of \textbf{CDP}. (a) An overview of the CDP model.
    (b) Architecture of Condition MoE Fusion. To fusion personalized contextual condition.
    (c) The pipeline of guided interest genatation. }
    \label{fig:framework}
\end{figure*}

\subsection{Overview}
Our proposed framework, \textbf{Contextual Diffusion Purifier (CDP)}, comprises three sequential stages: (1)~\textbf{Category-Gated Sequence Pooling and Noise Injection}; (2)~\textbf{Adaptive Condition Construction with Mixture-of-Experts}; and (3)~\textbf{Conditional Denoising and Downstream Embedding}. Figure~\ref{fig:framework} illustrates the complete pipeline.

\subsection{Problem Formulation}
Let $\mathcal{U}$ and $\mathcal{I}$ denote the user and item spaces, respectively.  
For a target interaction tuple $(q,u,i,s)$ where  
$q\in\mathcal{Q}$ is the query,  
$u\in\mathcal{U}$ the user,  
$i\in\mathcal{I}$ the candidate target item, and  
$s\in\mathcal{S}$ the contextual meta-data (hour, page, client, \emph{etc}.),  
our goal is to learn a context-adaptive interest embedding $\mathbf{z}^*\in\mathbb{R}^d$ such that  
\begin{equation}
p(\text{click}\mid q,u,i,s) = \sigma\bigl(\,f_\text{DNN}\bigl(\mathbf{z}^* \,\|\, \phi_\text{sparse}(q,u,i,s)\bigr)\bigr),
\end{equation}
where $\sigma(\cdot)$ is the sigmoid function and $\phi_\text{sparse}(\cdot)$ the original sparse features.  
We obtain $\mathbf{z}^*$ by denoising the observed behavior sequence under the joint context.

\subsection{Category-Gated Sequence Pooling and Noise Injection}
\label{subsec:pool_noise}

Given the raw user sequence $\mathcal{H}=\{e_1,\dots,e_T\}$ with $e_t\in\mathbb{R}^d$, we first filter out cross-category noise by retaining only items whose category matches that of the target item $i$:
\begin{equation}
\mathcal{H}_\text{cat} = \bigl\{e_t\in\mathcal{H}\mid \text{category}(e_t)=\text{category}(i)\bigr\}.
\end{equation}
The filtered sequence is then mean-pooled into an initial intent vector
\begin{equation}
\mathbf{z}_0 = \frac{1}{|\mathcal{H}_\text{cat}|}\sum_{e_t\in\mathcal{H}_\text{cat}} e_t.
\end{equation}
Critically, the aggregation of user behavior sequences operates as a low-pass filter, attenuating high-frequency behavioral interest components through inherent noise introduction. Ulteriorly, to model the corruption process, we follow the standard diffusion schedule and inject Gaussian noise at step $t$:
\begin{equation}
\mathbf{z}_t = \sqrt{\bar{\alpha}_t}\,\mathbf{z}_0 + \sqrt{1-\bar{\alpha}_t}\,\varepsilon,\quad \varepsilon\sim\mathcal{N}(0,\mathbf{I}),
\label{eq:forward}
\end{equation}
where $\bar{\alpha}_t=\prod_{k=1}^t\alpha_k$ is the cumulative product of noise schedule coefficients.

\subsection{Adaptive Condition Construction}
\label{subsec:condition}

To extract comprehensive contextual signals for user interest generation, we construct three complementary conditions ( Collaborative Condition, Contextual \& Scene Conditions) and adaptively fuse them via a lightweight Mixture-of-Experts (MoE) framework, enabling personalized contextual control.

\paragraph{Collaborative Condition}
The collaborative condition captures the triadic interaction among query, user, and item.  
We first concatenate the raw embeddings and their bilinear interactions:
\begin{equation}
\mathbf{c}_\text{co} = \phi_\text{co}\Bigl(\bigl[\mathbf{q}\;\|\;\mathbf{u}\;\|\;\mathbf{i}\;\|\;\mathbf{q}\!\odot\!\mathbf{u}\;\|\;\mathbf{u}\!-\!\mathbf{i}\;\|\;\mathbf{q}\!\odot\!\mathbf{i}\;\|\;\mathbf{q}\!-\!\mathbf{i}\bigr]\Bigr),
\end{equation}
where $\phi_\text{co}$ is a two-layer MLP with ReLU activations.

\paragraph{Contextual \& Scene Conditions}
Contextual features $\mathbf{x}_\text{ctx}$ (hour-of-day, page index, \emph{etc}.) and scene features $\mathbf{x}_\text{scene}$ (client type, search\_source, \emph{etc}.) are separately encoded via two-layer MLPs $\phi_\text{ctx}$ and $\phi_\text{scene}$:
\begin{equation}
\mathbf{c}_\text{ctx} = \phi_\text{ctx}(\mathbf{x}_\text{ctx}), \qquad
\mathbf{c}_\text{scene} = \phi_\text{scene}(\mathbf{x}_\text{scene}).
\end{equation}

\paragraph{MoE Fusion}
 To obtain adaptive personalized guidance condition, the three conditions are concatenated and fed into a gating network:
\begin{equation}
\mathbf{g} = \operatorname{softmax}\bigl(\mathbf{W}\,[\mathbf{c}_\text{co}\;\|\;\mathbf{c}_\text{ctx}\;\|\;\mathbf{c}_\text{scene}]\bigr),
\end{equation}
where $\mathbf{W}\in\mathbb{R}^{3\times 3d}$.  
The final condition vector is the gated sum:
\begin{equation}
\mathbf{c} = \sum_{j\in\{\text{co,ctx,scene}\}} g_j\,\mathbf{c}_j.
\end{equation}

\subsection{Conditional Denoising Network}
\label{subsec:denoise}
As established in our formal framework, the reverse diffusion phase employs a multi-layer MLP denoising network as denoiser $\epsilon_\theta(\mathbf{z}_t,t,\mathbf{c})$ to perform conditioned iterative denoising of user behavior Gaussian noise $\mathbf{z}_t$ into purified user-adaptive interest representations, guided by contextual signal $\mathbf{c}$.

Formally, the denoise process at diffusion reverse step $t$ is
\begin{equation}
\mathbf{h}_{t+1} = \epsilon_\theta(\mathbf{h}_t,t,\mathbf{c}),
\end{equation}
where $\mathbf{h}_{0} = \mathbf{z}_t$.

The final prediction is
\begin{equation}
\hat{\varepsilon} = \mathbf{W}_\text{out}\,\mathbf{h}_L,
\end{equation}
where $L$ is the total denoise step, and the training objective is jointly constrained by standard noise-prediction loss $\mathcal{L}_\text{recon}$ and cross-entropy loss $\mathcal{L}_\text{BCE}$:
\begin{align}
\mathcal{L}_\text{loss} &= \mathcal{L}_\text{recon} + \mathcal{L}_\text{BCE}
\\\\ &= \mathbb{E}_{\mathbf{z}_0,\varepsilon,t}\Bigl[\bigl\|\hat{\varepsilon}-\varepsilon\bigr\|_2^{2}\Bigr] + ylog(\sigma)+(1-y)log(1-\sigma)
\end{align}
where $\sigma$ is the final predict score and $y$ is the label.

\subsection{Inference and Downstream Integration}
\label{subsec:inference}

During online inference, we execute a 20-step denoising process on user interest representations $\mathbf{z}_t$ to achieve an optimal efficiency-effectiveness trade-off. The resulting context-adaptive user embeddings seamlessly integrate with downstream interaction modules without architectural modifications, while incurring only marginal latency overhead.

\section{Experiments}

\subsection{Experiment Setup}
\subsubsection{Datasets}
Our model is based on user behavior sequence modeling using a search system. Therefore, in order to conduct fair testing, the dataset needs to include the following three features: 1) rich user behavior with sufficiently dense signals for generating user interests, 2) user search queries, and 3) sufficient contextual information with a complete conditional space to guide user interest generation. However, the existing dataset cannot simultaneously meet these three requirements. For this consideration, we collected real user logs based on online platforms as the training and testing sets. Specifically, the full user behavior of the first 60 days was used for training, and the user behavior of the last day was used for testing. The sample size for a single day is 300 million, with an average user behavior sequence length of 2000 and rich user contextual information, providing a large-scale and realistic basis for evaluating our conditional diffusion behavior model.

\subsubsection{Metrics}

In our experiments, we adopt standard evaluation metrics for ranking models: AUC, User AUC (UAUC), and Group AUC (GAUC). UAUC assesses ranking quality by computing AUC per user, while GAUC evaluates performance per request group, making it a more accurate proxy for online traffic patterns.

\subsubsection{Parameter Setting}

We train our model using the Adam~\cite{adam2014method} optimizer. The training task is conducted on four NVIDIA A10 GPUs, with each GPU handling a batch size of 4096. The learning rate is set to 0.00025 for dense parameters and 0.0005 for sparse parameters during the training process. The diffusion process is carried out in 100 steps for both noise addition and denoising. A linear sampling schedule is used, with a noise intensity ranging from 0.005 to 0.01.

\subsubsection{Baselines}

We compare our method with two type methods: User behavior based:  , SIM ~\cite{10.1145/3340531.3412744}, QIN~\cite{10.1145/3583780.3615022}, TWIN~\cite{10.1145/3580305.3599922}, HSTU~\cite{10.5555/3692070.3694484}, and two diffusion based methods: DDRM~\cite{10.1145/3626772.3657825}, DiffRec~\cite{10.1145/3539618.3591663}.

\textbf{SIM}~\cite{10.1145/3340531.3412744} orchestrates ultra-long behavior sequences via a coarse-to-fine paradigm: GSU first prunes the search space to a tractable subset, after which ESU refines attention for precise modeling.

\textbf{QIN}~\cite{10.1145/3583780.3615022} boosts intent-aware ranking by mining query-relevant user history via a search unit, filtering interactions for query-aligned slices.

\textbf{TWIN}~\cite{10.1145/3580305.3599922} employs a shared attention mechanism to maintain consistency between GSU and ESU, enabling efficient processing of ultra-long behavior sequences.

\textbf{HSTU}~\cite{10.5555/3692070.3694484} employs an autoregressive architecture to model user–item interactions as sequential token generation, treating recommendation as a sequence-to-sequence transduction task.

\textbf{DDRM}~\cite{10.1145/3626772.3657825} utilizes a multi-step denoising process of diffusion models to enhance the robustness of user and target item embeddings, boosting recommendation performance.

\textbf{DiffRec}~\cite{10.1145/3539618.3591663} generates personalized user representations by adding and removing noise from interaction histories, accurately modeling user behavior and preferences.

We implemented baseline methods using configurations specified in their original papers, maintaining comparable parameter scales and identical training durations to CDP for fair experimental comparison.

\subsection{Experimental Results}

\begin{table}[htbp]
\centering
\caption{Overall performance comparison with state-of-the-art methods}
\label{tab: overall performance}
\begin{tabular}{cccc}
\hline
Model   & AUC & UAUC & GAUC \\ \hline
SIM~\cite{10.1145/3340531.3412744}     &0.7298     &0.6425      & 0.6283     \\
QIN~\cite{10.1145/3583780.3615022}    & 0.7314    &0.6437      & 0.6301     \\
TWIN~\cite{10.1145/3580305.3599922}     &0.7329     &0.6451      & 0.6312     \\
HSTU~\cite{10.5555/3692070.3694484}    &0.7332     & 0.6455     & 0.6310     \\
DDRM~\cite{10.1145/3626772.3657825}    &0.7321     &0.6429      &0.6295      \\
DiffRec~\cite{10.1145/3539618.3591663} &0.7327     &0.6447      &0.6315      \\
\textbf{CDP}     &\textbf{0.7402}     &\textbf{0.6514}      & \textbf{0.6362}     \\ \hline
\end{tabular}
\end{table}

We first compare our proposed CDP with four representative user behavior modeling baseline methods on CTR prediction tasks. As shown in Table~\ref{tab: overall performance}, CDP consistently achieves the best performance on all metrics. Specifically, CDP obtains the highest AUC (0.7402), UAUC (0.6514) and GAUC (0.6362), outperforming all baselines by a significant margin. In addition, compared with the two diffusion-based models, our proposed CDP also shows significant improvement. These results indicate that our proposed model can generate more personalized adaptive representations based on contextual conditions, achieving superior results.

\subsection{Ablation Study}
To validate the efficacy of our multi-condition fusion mechanism, we conducted systematic ablation studies, comparing CDP against four ablated variants: (i) no conditions, (ii) no collaborative condition, (iii) no context-and-scene conditions, and (iv) no MoE fusion. As illustrated in Table \ref{tab: Ablation}, CDP consistently outperforms all ablations, underscoring the indispensability of each design choice. Removing the MoE fusion cripples the model’s ability to adaptively balance conditions, yielding the largest drop. Moreover, the collaborative condition emerges as the most influential cue—an outcome fully aligned with our intuition.

\begin{table}[]
\centering
\caption{Ablation Study of CDP.}
\label{tab: Ablation}
\begin{tabular}{cccc}
\hline
Model                                                                & AUC    & UAUC   & GAUC   \\ \hline
No condition                                                         & 0.7341 & 0.6467 & 0.6336 \\
No collaborative & 0.7362 & 0.6475 & 0.6348 \\
No other conditions                                                  & 0.7382 & 0.6492 & 0.6355 \\
No MoE fusion                                                        & 0.7355 & 0.6470 & 0.6341 \\
\textbf{CDP}                                                                & \textbf{0.7402} & \textbf{0.6514} & \textbf{0.6362} \\ \hline
\end{tabular}
\end{table}

\subsection{Online A/B test}
We have deployed our proposed model in a onine E-Commerce Search Platform to evaluate the online effectiveness. Our online base model is similar to the SIM-based model. We conducted an online experiment for 14 days, involving billions of user requests, and observed improvements in four main metrics, as shown in Table~\ref{tab: online}. The metrics of \#Total Orders and \#GMV (Gross Merchandise Value) measure the user engagement on our online shopping platform, while online CTR measure the effectiveness of the models. Online experiments reveal the superior performance of CDP in real-world applications, providing a better shopping experience for billions of users with acceptable latency (+2ms).

\begin{table}[htbp]
  \centering
  \caption{Online A/B test metric of CDP.}
  \begin{tabular}{lcccc}
    \toprule
    Method & \#Total GMV & \#Total Orders & CTR & latency \\
    \midrule
    Base & - & - & - & -\\
    Proposed & \textbf{+1.240\%} & \textbf{+1.032\%} & \textbf{+0.745\%}  & \textbf{+2ms} \\
    \bottomrule
  \end{tabular}
  \label{tab: online}
\end{table}

\section{Conclusion}
In this paper, we proposed the Contextual Diffusion Purifier (CDP), a novel framework that decouples noise from genuine interests in behavioral sequences via query-user-item-context conditioned diffusion. By treating behaviors as contaminated observations and adaptively fusing multi-source conditions through MoE, CDP generates context-aware representations that dynamically purify signals from exposure bias and category drift. Compared with the recent state-of-the-art methods, extensive offline evaluations on industrial datasets confirm the effectiveness of our approach, and large-scale online A/B tests on real-world traffic have delivered significant gains—underscoring its real-world practicality.

Despite CDP's robust performance, its efficacy may be compromised in scenarios with low behavioral density—particularly for cold-start and low-activity users where insufficient signals introduce modeling bias. Concurrently, our current framework employs coarse-grained condition selection and fusion mechanisms, limiting precise interest control. To address these limitations, future work will augment behavioral representations through meta-learning and cross-domain signal transfer techniques while developing fine-grained condition extraction methods to construct comprehensive conditional embeddings, enabling stronger controlled generation of context-aware user interests.

\bibliographystyle{ACM-Reference-Format}
\bibliography{reference}
\end{document}